# Network approach reveals the spatiotemporal influence of traffic to air pollution under the COVID-19

Weiping Wang[1], Saini Yang[1,2], Kai Yin[3], Zhidan Zhao[4], Na Ying[5,*], Jingfang Fan[6,*]
1 School of National Safety and Emergency Management, Beijing Normal University, Zhuhai 519087, China
2 State Key Laboratory of Earth Surface Processes and Resources Ecology, Beijing Normal University, Beijing 100875, China
3 School of Traffic and Transportation, Beijing Jiaotong University, Beijing 100044, China
4 China Complexity Computation Lab, Department of Computer Science, School of Engineering, Shantou University, Shantou 515063, China
5 China State Key Laboratory of Environmental Criteria and Risk Assessment, Chinese Research Academy of Environmental Sciences, Beijing, 100012, China.
6 School of Systems Science, Beijing Normal University, Beijing 100875, China
*Corresponding author: yingna@mail.bnu.edu.cn; jingfang@bnu.edu.cn

**Abstract**

Air pollution causes widespread environmental and health problems and severely hinders the life quality of urban residents. Traffic is a critical for human life and its emissions are a major source of pollution, aggravating urban air pollution. However, the complex interaction between the traffic emissions and the air pollution in the cities has not yet been revealed. In particular, the spread of the COVID-19 has caused various cities to implement different traffic restriction policies according to the local epidemic situation, which provides the possibility to explore the relationship between urban traffic and air pollution. Here we explore the influence of traffic to air pollution by reconstructing a multi-layer complex network base on traffic index and air quality index. We uncover that air quality in Beijing-Tianjin-Hebei (BTH), Chengdu-Chongqing Economic Circle (CCS) and Central China (CC) regions are significantly influenced by the surrounding traffic conditions after the outbreak. Under different fights against the epidemic stages, the influence of traffic in other cities on the air pollution reached the maximum in stage 2 (also called Initial Progress in Containing the Virus). For BTH and CC regions, the impact of traffic on air quality becomes larger in the first two stages and then decreases, while for CC, the significant impact occurs in Phase 3 among regions. For other regions, however, the changes are not evident. Our presented network-based framework provides a new perspective in the field of transportation and environment, and maybe helpful to guide the government to formulate air pollution mitigation and traffic restriction policies.

**The rising of air pollution has significant negative impact on human health. Urban traffic emissions will increase air pollution. However, their dynamic modes among cities have still remained a challenging problem. The impact of the epidemic has led cities to implement different traffic restriction policies one after another, which naturally forms a controlled experiment to reveal their relationship. In the present work, we develop a multi-layer network-based framework with traffic index and air quality index. We find that air quality is related to their surrounding traffic condition in BTH, CCS and CC when the epidemic spreads. In addition, the different fight against the epidemic stages of influences of traffic on SAT are identified. Our method and results presented here not only provide the deep understanding of the influence of traffic to air pollution, but also can be applied to study other climate and environment phenomena, such as the global warming.**

## I. INTRODUCTION

With the use of fossil energy sources such as transportation, industry, agriculture, power generation, as well as the daily cooking and heating needs of residents continue to increase, eventually leads to continuous serious air pollution problems. Air pollution poses a major threat to human health and can cause stroke, heart disease, lung cancer, acute and chronic respiratory diseases[1]. According to WHO estimates, about 91% of the world's population lives in places with poor air quality, causing about 4.2 million deaths per annum[2]. Transportation system is one type of critical lifeline and is essential to the operation of modern society[3]. Traffic emission is a major source of air pollution in urban areas[4,5], including carbon monoxide (CO), carbon dioxide ($CO_2$), volatile organic compounds (VOCs) or hydrocarbons (HCs), nitrogen oxides (NOx), secondary aerosols formed through physical and chemical processes and pollutants-suspended particles caused by brakes, tire wear and re-wear[6]. Urban traffic congestion have increased traffic emissions, and further increases air pollution. Thus, it is vital to explore the relationship of traffic to air pollution.

Considerable research has been devoted to exploring the influence of traffic to air pollution. The most intuitive method is through spatiotemporal variations of air pollutants including CO, CO2, PM2.5, PM10, SO2, O3 and NOx. With diurnal analysis of hourly PM2.5, PM10, NO2, and CO concentrations, the two ascending stages caused by the two traffic peaks were found[7]. An increase in secondary organics (NH4)2SO4 and NH4NO3 caused by vehicle emission were also found[8]. Then, some statistical methods are used in this field. multivariate autoregressive models are used to estimate pollution levels under different traffic conditions[9]. Further, some physical approaches are proposed based on a parameterized analytical representation of the entire fuel consumption and emissions process. These approaches like CMEM[10] and MOBILE[11] can simulate and estimate the CO, HC, NOx emitted by traffic[12]. Experimental studies are also exploring the impact of traffic and its emissions on air pollution. Combining the chassis dynamometer system and an outdoor enclosed environmental chamber, new particle formation from traffic emissions has been assessed. The new particle formation can frequently produce high levels of ultrafine particles, causing serious air pollution[13]. However, the above methods are either microscopically unable to explore the relationship between traffic and air pollution in large-scale areas (like physical and experimental methods), or generalize and fail to reveal the mechanism of how regional traffic variability causes the change of air pollution (like statistical methods).

With the corona virus pandemic sweeping the world, many countries have implemented strict lockdown policies to stop the spread of the disease. These policies, especially limited transportation activities, will improve ambient air quality. This has been confirmed in some cities or regions in China[4,5,21–24,6,14–20], Egypt[25], Spain[26], France[26], Italy[26–28], Brazil[29–31], Korea[32], New Zealand[33], Singapore[34], the United States[9,26,35], Malaysia[36], East Asia[37], Europe[38] and even at the global level[39]. However, the study[40] has also found that the overall air quality in urban areas in China has not improved with the COVID-19 lockdown. Besides, as the world gradually opened up due to the relaxation of movement restrictions, the harsh air returns to some city in Asia. For example, India is worst now than before[41]. With the lifting of the lockdown in some areas of China, large-scale movements of people and goods began, air pollution gradually returns to or likely to exceed the level before the quarantine[42]. Therefore, with the different development stages and response measures of the pandemic, the contribution of traffic to air pollution will also change significantly. An improved understanding the role of traffic in air pollution during the spatiotemporal changes of the COVID-19 is thus needed. Since it is beneficial to reduce air emissions through regulation and incentives.

During the past years, network theory has been found useful for better understanding spatiotemporal behavior in the climate system[20-22]. Climate networks establish correlations among climate anomalies in distant parts of the world and attempt to explain them using relevant physical progress. In a climate network, geography data are transformed into nodes and edges of a network that can represent spatiotemporal relationships. Nodes refer to geographical locations or grid sites, and edges are constructed based on similarities (such as cross-correlations) in the variability over time between pairs of nodes. Various climate data records (such as temperature, pressure, winds, and precipitation) can be used to construct a climate network. Climate network approach can provide a powerful framework to better understanding the structure and pattern of climate phenomena including air pollution[43].

The basic idea behind climate networks is that relevant and important features of atmospheric mechanisms influence the variability of traffic index to air pollution at different locations, and these influences are encoded in the structure of the network. By extracting the topological index of the network, we can reveal the underlying links from traffic to air pollution. In this study, we develop a network-based framework to explore the influence of traffic on air pollution during the temporal and spatial changes of the COVID-19. A multi-layer network between traffic index and air pollution is reconstructed. Our results can help to formulate strategies and countermeasures for traffic emission and air pollution.

## II. DATA

### A. traffic index data.

In this study, we collect the traffic index data from TOMTOM (https://www.tomtom.com/en_gb/traffic-index/). These data can represent congestion levels in Chinese cities. Here we employ daily traffic index (TL) of 21 major cities from January 1 to July for 2019 and 2020, respectively. TL is calculated by calculating the proportion of increase in actual travel time over free flow travel time and its value is greater than or equal to 0. The larger the indicator value, the more serious the traffic congestion is.

### B. Air quality index

Daily air quality index (AQI) index data acquired from the China National Environmental Monitoring Centre (CNEMC) are used in this study. AQI is based on ambient air quality standards and the impact of various pollutants on human health, ecology and the environment, and simplifies the concentrations of several air pollutants that are routinely monitored into a single index value. The value range of AQI is set from 0 to 500. The larger the value, the more serious the air pollution. According to the traffic indexes (TL) records, AQI index of 21 major cities from January 1 to July for 2019 and 2020 are selected.

## III. METHODS

### A. Data pre-processing

In this study, we divided 21 cities into six region groups according to the existing geographical divisions namely Beijing-Tianjin-Hebei (BTH), Northeast China (NEC), Chengdu-Chongqing Economic Circle (CCS), Central China (CC), Guangdong-Hong Kong-Macao Greater Bay Area (GHM), and Yangtze River Delta (YRD) as shown in Table 1. The BTH includes two municipalities, Beijing and Tianjin and is located in the heart of the Bohai Rim in China. It is the largest and most dynamic region in northern China, one of the regions with the greatest potential for economic development in China, and one of the regions with the most intensive transportation and logistics network. The one-hour traffic circle with rail transit has been initially formed. NEC is the general term for the land in the northeast of China. The economy of the NEC started early and has made great historical contributions to the development and growth of China. However, due to the serious loss of young people and the impact of cold weather, the economic slowdown of NEC has caused it to fail to keep up with the pace of the whole country in the past 30 years, and it has also affected the construction of transportation infrastructure. The CCS is an urbanization area with the highest level of development and great development potential in western China. It is an important part of the implementation of the Yangtze River Economic Belt and the Belt and Road strategy. It is the starting point of the new land-sea corridor in the west, and has the unique advantage of linking the southwest and northwest, and connecting East Asia with Southeast Asia and South Asia. CC is located in the central part of China, with many national transportation trunk lines reaching the whole country. It has the advantage of being a strategic hub in the east, west, north and south of the country and a water and land transportation hub. Economically, it is considered to be a relatively underdeveloped area. The GHM includes Hong Kong, Macau, Guangzhou, Shenzhen and other cities. The GHM, the New York Bay Area, the San Francisco Bay Area, and the Tokyo Bay Area of Japan are also known as the four major bay areas in the world. It is one of the regions with the highest degree of openness and the strongest economic vitality in China. It has an important strategic position in the overall development of the country and a convenient and efficient modern comprehensive transportation system is being formed rapidly. The YRD is an important intersection between the "Belt and Road Initiative" and the Yangtze River Economic Belt. It is an important platform for China to participate in international competition, an important engine for economic and social development, and one of the regions with the best urbanization foundation in China. The density of highway and railway transportation lines in the YRD leads the country, and a three-dimensional comprehensive transportation network is basically formed.

By using the cumulative confirmed cases of cities as of March 16, 2021 to represent risk of COVID 19, we can classified cities into 4 outbreak levels: 1:[0,100), 2:[100, 300), 3:[300, 1000) and 4: [1000,+ ∞) as shown in Table 1. The higher outbreak level of the city, the higher the number of infected people and the wider the spread of the epidemic in the city. Here, the effect of seasonality in AQI has been removed by subtracting the calendar day's mean from the original datasets.

**Table 1 Outbreak level for COVID-19 among cities**

| City | Outbreak level | Region group | Cumulative confirmed cases |
|---|---|---|---|
| Beijing | 4 | BTH | 1049 |
| Tianjin | 3 | BTH | 364 |
| Shijiazhuang | 3 | BTH | 898 |
| Shenyang | 1 | NEC | 70 |
| Changchun | 2 | NEC | 150 |
| Chengdu | 2 | CCS | 158 |
| Chongqing | 3 | CCS | 591 |
| Wuhan | 4 | CC | 50340 |
| Changsha | 3 | CC | 242 |
| Guangzhou | 3 | GHM | 377 |
| Shenzhen | 3 | GHM | 423 |
| Zhuhai | 1 | GHM | 98 |
| Dongguan | 1 | GHM | 99 |
| Xiamen | 1 | GHM | 35 |
| Quanzhou | 1 | YRD | 47 |
| Shanghai | 4 | YRD | 1840 |
| Suzhou | 1 | YRD | 87 |
| Wuxi | 1 | YRD | 55 |
| Nanjing | 1 | YRD | 93 |

| Hangzhou | 2 | YRD | 181 |
| Ningbo | 2 | YRD | 157 |

## B. Network construction

Similar to earlier studies[44,45], we define the $X_{T_j,A_i}(\tau)$ as the time-delayed cross-correlation function for the TL node *j* and AQI node *i*, denoted by $X_{T_j,A_i}(\tau)$: for $\tau \geq 0$,

$$X_{T_j,A_i}(\tau) = \frac{\sum_{t=1}^{L-\tau}(A_i(t)-\bar{A}_i)(T_j(t+\tau)-\bar{T}_j)}{\sqrt{\sum_{t=1}^{L-\tau}(A_i(t)-\bar{A}_i)^2} \cdot \sqrt{\sum_{t=1}^{L-\tau}(T_j(t+\tau)-\bar{T}_j)^2}} \quad (1)$$

and for $\tau < 0$,

$$X_{T_j,A_i}(-\tau) = \frac{\sum_{t=1}^{L-\tau}(A_i(t+\tau)-\bar{A}_i)(T_j(t)-\bar{T}_j)}{\sqrt{\sum_{t=1}^{L-\tau}(A_i(t+\tau)-\bar{A}_i)^2} \cdot \sqrt{\sum_{t=1}^{L-\tau}(T_j(t)-\bar{T}_j)^2}} \quad (2)$$

where $\bar{A}_i$ and $\bar{T}_j$ denotes the average of AQI time series and TL time series. The time lags $\tau$ are in the range between −7 and +7 days. The time lag is chosen to be long enough to avoid the sensitive of correlation estimation to our choice of time lag which leads to erroneous correlation estimation. The deviations in the link identification due to persistence or autocorrelation in the records are reduced through dividing the $std(X_{j,i})$. The strength of the positive and negative link weights is denoted as:

$$W_{T_j,A_i}^{pos} = \frac{(max(X_{T_j,A_i}) - mean(X_{T_j,A_i}))}{std(X_{T_j,A_i})} \quad (3)$$

$$W_{A_i,T_j}^{neg} = \frac{(min(X_{T_j,A_i}) - mean(X_{T_j,A_i}))}{std(X_{T_j,A_i})} \quad (4)$$

Where $max(X_{T_j,A_i})$, $min(X_{T_j,A_i})$, $mean(X_{T_j,A_i})$, and $std(X_{T_j,A_i})$ are the maximum, the minimum, mean, and the standard deviation of the cross-correlation function, respectively. We define $\tau_{T_j,A_i}^{pos}$ and $\tau_{T_j,A_i}^{neg}$ as the corresponding time lags at these two peaks. When $\tau_{T_j,A_i} > 0$, the links are outgoing from TI nodes pointing to AQI nodes; when $\tau_{T_j,A_i} < 0$, the links are pointing away from AQI nodes incoming to TI nodes. Here links with zero-time lags are excluded. The adjacency matrix of a climate network is defined as

$$\Lambda_{T_j,A_i}^{pos} = \begin{cases} 1 & if\ W_{T_j,A_i}^{pos} \geq Q\ and\ \tau_{T_j,A_i}^{pos} > 0 \\ 0 & else \end{cases} \quad (5)$$

$$\Lambda_{T_j,A_i}^{neg} = \begin{cases} 1 & if\ W_{T_j,A_i}^{neg} \leq -Q\ and\ \tau_{T_j,A_i}^{neg} > 0 \\ 0 & else \end{cases} \quad (6)$$

Here *Q* is a threshold for the weight links, which is determined based on the shuffling procedure[46,47]. In the shuffled case, the order of days is permutated for each pair of TI and AQI nodes *j* and *i*[47]. By this step we keep all the statistical quantities of the original data but omit the physical dependencies between TI and AQI nodes. In such a case, the shuffled network represents the properties of statistical quantities and the autocorrelations of the original records, which may introduce unrealistic links. If the original link weights are significantly higher than that of the control, we regard it as a real link; otherwise, they are spurious links. Then, we obtain the desired connection between TI and AQI based on the adjacency matrix $\Lambda_{T_j,A_i}^{pos}$ and $\Lambda_{T_j,A_i}^{neg}$.

The degree is the most common application for measuring climate networks. A link that points toward a node is referred as an in-degree link, and a link that points away from a node is considered as an out-degree link. The way of TI dynamically influenced to AQI is defined as weighted out-degree of TI nodes, which are the total outgoing link weights from TI nodes[48]. And the response of AQI to TI is denoted as weighted in-degree of AQI nodes, which are the total incoming links weights pointing towards to AQI nodes.

Obviously, the outgoing links of the TI are the same as the incoming links of the AQI. Nodes have higher values represent a larger amount connection with other nodes in the network; while lower ones mean "isolated" in the network. The In and Out fields describe the level of TI nodes impact on the AQI nodes and the level of affected AQI node from TI nodes, respectively.

## C. Significance tests

The statistical significance of link weights is determined based on a shuffling procedure. In the shuffled case, the order of years is permutated and the order of days within each year is maintained for each pair of TI and AQI nodes *i* and *j*[41]. We generate shuffled data according to the procedures described in the previous subsection. This shuffling keeps the all the statistical quantities of the original data but omits the physical dependencies between TI and AQI nodes. In such a case, the shuffled network represents the properties of statistical quantities and the autocorrelations of the original records, which may introduce unrealistic links. We choose a control for the records to distinguish realistic links from unrealistic ones. If the original link weights are significantly higher than that of the control, we regard it as a real link; otherwise, they are spurious links.

## D. Analysis Framework

In this study, we use two types of networks: single and two-layered networks and four steps to explore the influence of traffic to air pollution during the pandemic as shown in Figure 1. Specifically, we take the city as a node, and first construct a single-layer network of air quality in 2019 and 2020 to compare and study the changes in air quality during the epidemic. Then, some multi-layer networks between TL and AQI are constructed to explore the impact of the epidemic on air pollution, specifically, the impact of different regions (six regions), different time stages (five stages) and outbreak levels (four levels) of the development of the epidemic.

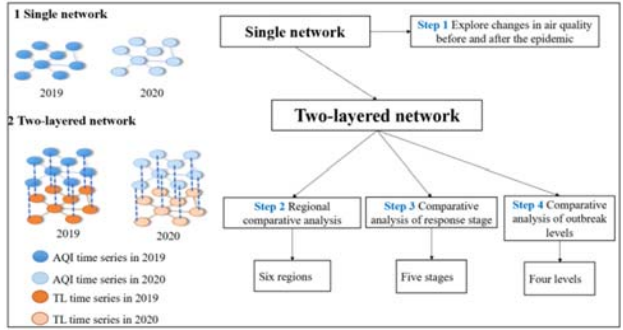

**FIG. 1** The analysis framework.

## IV. RESULTS AND DISCUSSION

We present the main results of the correlated multi-layered networks composed of the TI and AQI as described above.

**The change of air pollution with pandemic.**

We explore the change of air pollution with pandemic. The influence of target city' air pollution from other city is quantified by the weighted degrees associated with the total weights of the significant interlinks from other cities' nodes, which are presented in Figure 2. A higher weighted in-degree (WID) indicates that target cities receive haze from other cities, whereas a higher weighted out-degree denotes the stronger transport strength from target cities to other cities.

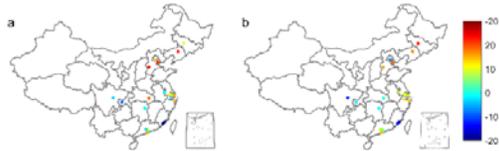

**FIG. 2** The maps of difference of weighted in- degrees (outgoing from the AQI nodes) between 2019 and 2020 (a) and the maps of difference of weighted out- degrees (incoming to the AQI nodes) between 2019 and 2020 (b).

We find that the value of weighted indegrees in the Beijing-Tianjin-Hebei (BTH) and Yangtze River Delta (YRD) regions have higher values in 2019. Compared with before the epidemic (2019), Beijing, Tianjin, Shijiazhuang, and Changchun are all decreasing in North China, especially in Beijing. In the YRD region, Nanjing, Hangzhou, and Suzhou are increasing, and other cities remain unchanged. Wuhan is decreasing, while Changsha City remains the same in Central China (CC). In the southwest China (CCS), Chengdu is decreasing, Chongqing is slightly intensified. For the Guangdong-Hong Kong-Macao Greater Bay Area (GHM), Guangzhou, Dongguan, Shenzhen, and Zhuhai are all increasing, while Quanzhou and Xiamen basically remain unchanged.

In terms of outdegree, the values of Nanjing and Hangzhou in the southwest region, Fujian Province and the YRD region are higher. Compared with before the epidemic (2019) in northern China, the value of Beijing has become larger; the value of Changchun has decreased and others have remained basically unchanged. In the YRD, the value of Hangzhou has become significantly smaller. In central China, the value of Wuhan has become larger. The southwestern region basically remained unchanged. Among the GHM, only Guangzhou was significantly smaller.

That is since some large cities are affected by the pandemic, their own production and the traffic are restricted. The contribution of air pollution from other cities has become greater.

Overall, we uncover those large cities, such as Beijing, Tianjin, and Wuhan have larger indegree values and lower outdegree values. This finding indicates that the air condition over these cities has less relative to air condition in the cities around it. In contrast, air quality levels in Hangzhou and Guangzhou have a weak relationship with their surrounding cities, hence, they are more likely influenced by the epidemic. In addition, there are no distinct changes for cities over most of the CES and GHM regions, suggesting that air quality over these cities is less influenced by the epidemic.

**The influence of traffic to air pollution with pandemic.**

Based on the weighted degree index, we further study the influence of traffic to air pollution with pandemic in different regions by constructing a multi-layer network between TL and AQI. The Figure 3 illustrates violin plots of the weighted in-degrees (outgoing from the TL nodes) among different regions. After comparing 2019 and 2020 in Figure 2, we found that the pandemic causes significant changes of influence of traffic to air pollution. Specifically, when an epidemic occurs in BTH, the air pollution of the city is greatly affected by the traffic of other cities (the weight in degree value becomes larger). The influence of traffic from other cities to this city's air pollution has changed from dispersed to larger ones for all cities in BTH. The change of influence of traffic to air pollution with pandemic in NEC and CCS regions is similar to BTH. Among BTH, NEC and CCS, weighted in- degrees of CCS has changed the most from before the epidemic. The regions of CC and GHM have larger mean value, larger maximum and smaller minimum value of weight in degree when the epidemic occurs. This suggests that some cities in these areas have closed traffic or factories for a period of time after the epidemic, while other cities are basically unaffected. What is more surprising is that compared with the epidemic in YRD, traffic in other cities has less impact on its own city's traffic pollution. The results indicate that the epidemic has no major impact on the transportation, production and life in these area, or they recovered much fast.

Affected by the epidemic, values over BTH, CC and CCS regions tend to be concentrated, suggesting that the air condition over these cities are largely related to their surrounding traffic. For YRD and GHM regions, values tend to be dispersed. Which means that the air quality level over these cities is less influenced by traffic from other cities.

Hence, their own pollution is mainly caused by the emissions and dust from mobile sources. The air pollution in NEC regions is hardly affected by the epidemic.

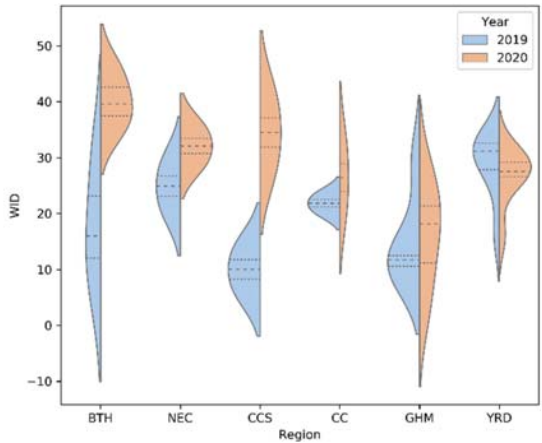

**FIG. 3**. Violin plots of the weighted in- degrees (WID) that are outgoing from the TL nodes among different regions.

**The influence of traffic to air pollution with different stages of fighting the pandemic.**

The epidemic situation and its control measures vary in different response stages, which will eventually lead to distinct impacts of traffic on air pollution. According the State Council Information Office of the People's Republic of China, China's fight against the epidemic of COVID-19 can be divided into five stages[49]: (I) Swift response to the public health emergency (December 27, 2019-January 19, 2020): the nationwide epidemic prevention and control plan was launched after cases is confirmed in Wuhan, as well as cases in other parts of China due to virus carriers traveling from the city; (II) Initial progress in containing the virus (January 20-February 20, 2020): The number of newly confirmed cases across the country is increasing rapidly, and the prevention and control situation is extremely severe. China adopted a key measure to stop the spread of the virus by closing outbound traffic from Wuhan and Hubei; (III) Newly confirmed domestic cases on the Chinese mainland drop to single digits (February 21-March 17, 2020): the epidemic prevention and control has achieved important results, people have resumed work and production in an orderly manner, urban traffic has resumed, and emissions have increased; (IV) Wuhan and Hubei – an initial victory in a critical battle (March 18-April 28, 2020): the spread of the local epidemic in the country with Wuhan as the main battlefield is basically blocked, and the control measures for the outbound traffic from Wuhan and Hubei are lifted; (V) Ongoing prevention and control (Since April 29, 2020): the domestic epidemic situation is generally sporadic, and there are clustered epidemics caused by sporadic cases in some areas. The national epidemic prevention and control has become normalized, and the traffic has returned to normal as a whole. To further study the relationship between traffic and air pollution, we classify our datasets into 5 groups by using time intervals of these five stages. The Figure 4 illustrates the the maps of weighted in- degrees for different stages.

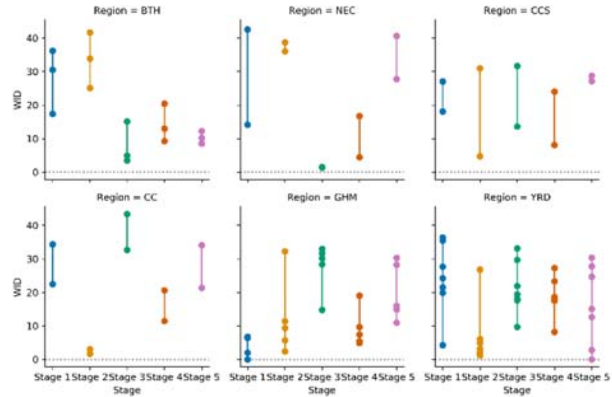

Fig. 4 The maps of weighted in- degrees, WID (incoming to the AQI nodes and outgoing from the TL nodes) for Stage I (a), II(b), III(c), IV(d) and V(e), respectively.

In Figure 4, we find that the impact of traffic in other cities on the city's air pollution reached the maximum in stage 2, and then reached the second peak in stage 4 in the region BTH. For region NEC, the impact reached the maximum in beginning stage, and then reached the second peak in the last stage. The impact of traffic reaches its maximum in Phase 3 among regions CCS and CC, especially in CC. For YRD, the traffic's impacts in each stage are basically the same.

From a different stage perspective in Figure 4, due to the spread of the epidemic in stage 1, travel in most cities is restricted, causing the pollution of this city to be greatly affected by the traffic of other cities, but the GHM region is still more affected by itself. Unlike other regions where the impact is high, air pollution in region CC is less affected by traffic from other cities in stage 2. As time goes on, the impact of traffic in other cities on air pollution in this city from region CC has changed from small to large in stage 3. In stage 4, the impact of traffic on air pollution in other cities is relatively small for all regions. At stage 5, only the traffic in other cities in region BTH has a relatively small impact on themselves city's air pollution.

Overall, the impact of traffic to air pollution in BTH and CC regions have a large fluctuation, while in CCS, GHM and YRD regions, the variations are much less. In Wuhan city, however, the values of weighted indegrees are consistent with variation during the epidemic.

**The influence of traffic to air pollution with different outbreak level of cities.**

Due to China's vast land, the development of the epidemic situation is significantly different. We need to deeply explore the relationship between traffic and air pollution in different epidemic development areas. According the COVID-19 cumulative confirmed cases as of March 16, 2021, we classified the level of outbreak level of cities (City level) into 4 groups as shown in Table I. The Figure 5 shows the weighted in- degrees (outgoing from the TL nodes) among different outbreak level of cities.

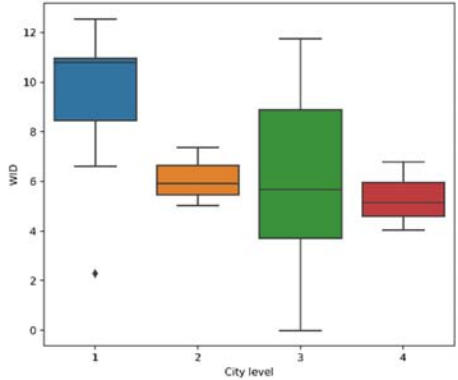

**Figure 5** Box plots of the weighted in- degrees (WID) that are outgoing from the TL nodes among different outbreak level of cities (City level).

From the average value of view, the weighted in-degree is mainly opposed to the outbreak level of the city. It shows that the greater of in-degree values, the smaller the outbreak level of the cities. When the outbreak level is high, the traffic in the target city and its surrounding areas is greatly affected by the epidemic, resulting in a less impact of traffic in surrounding cities on air pollution in the target city. Thus, traffic has little impact on the air quality of cities with high outbreak levels. Regarding the different outbreak levels of the level cities, the changes between level 1 and level 3 are relatively large, especially for level 3.

## V. CONCLUSIONS

In this study, both single layer and multi-layers networks have been developed to explore the influence of traffic on air pollution during the pandemic based on complex network approaches. We have found that the epidemic has less impact on air quality over Beijing, Tianjin, and Wuhan cities, while for Hangzhou and Guangzhou cities, their air quality is related to the epidemic. Compared with 2019, air quality in BTH, CCS and CC regions are connected with the surrounding traffic conditions. In contrast, there are no significant changes in YRD and GHM region between 2019 and 2020. Furthermore, we analyzed the variations during different epidemic stages. The impact of traffic in other cities on the city's air pollution reached the maximum in stage 2. For BTH and CC regions, the impact of traffic on air quality is large in the first two stages and then decreases trend, while for CC, the significant impact occurs in Phase 3 among regions. For other regions, there is little change in different stages. In addition, the impact over different outbreak level is also investigated. The higher outbreak level is generally having a lower indegree value. In the case of high ranking, the traffic of the surrounding cities has less impact on the air pollution of the target city.

Compared with traditional research methods, the climate network method used in this study can explore the relationship between traffic emissions and air pollution on a larger scale, especially the long-distance impact between different cities, and more macroscopically indicate the impact of other cities' emissions on this city' air pollution contribution. However, this study only uses teleconnection for network modeling, and uses the index of degree for quantitative analysis and lacked a more detailed and refined aerodynamic transmission mechanism, making the quantification of the contribution of traffic emissions to air pollution not precise enough. The division of urban outbreak levels and epidemic development stages are relatively subjective and lack more precise quantitative evaluation criteria. Subsequent research should choose a more scientific and effective division method. Besides providing information for guiding government policies to improve air quality levels, the network parameters in this work are a profitable attempt in the field of transportation and atmospheric environment. Our results also can call attention to further research on the impact of transportation on air pollution.

## ACKNOWLEDGMENTS

This study was supported by the National Natural Science Foundation of China No. 72001018, State Key Laboratory of Earth Surface Processes and Resource Ecology Open Fund (Grant No. 2020-KF-07). The participation of Dr. Na Ying in this study was supported in parts by the Budget Surplus of Central Financial Science and Technology Plan (Grant No. 2021-JY-15). The authors wish to thank TOMTOM for providing the traffic index datasets.

## DATA AVAILABILITY

The data that support the findings of this study are available from the corresponding author upon reasonable request.